\documentclass[prl,aps,showpacs,twocolumn,superscriptaddress,amsmath,amssymb,floatfix]{revtex4}
\usepackage{graphicx}
\usepackage{color}

\begin{document}
\title{Is there a no-go theorem for  superradiant quantum phase transitions \\  in
 cavity and circuit QED ?}

\author{Pierre Nataf}
\author{Cristiano Ciuti}
\email[E-mail: ]{cristiano.ciuti@univ-paris-diderot.fr}
\affiliation{Laboratoire Mat\'eriaux et Ph\'enom\`enes Quantiques,
Universit\'e Paris Diderot-Paris 7 and CNRS, \\ B\^atiment Condorcet, 10 rue
Alice Domont et L\'eonie Duquet, 75205 Paris Cedex 13, France}
\affiliation{}

\begin{abstract}
In cavity quantum electrodynamics (QED),  the interaction between an atomic transition and the cavity field is measured by the vacuum Rabi frequency $\Omega_0$. The analogous term "circuit QED" has been introduced for Josephson junctions, because superconducting circuits behave as artificial atoms coupled to the bosonic field of a resonator.  In the regime with $\Omega_0$ comparable to the two-level transition frequency, "superradiant" quantum phase transitions for the cavity vacuum have been predicted, e.g.  within the Dicke model. Here, we prove that if the time-independent light-matter Hamiltonian is considered, a superradiant quantum critical point is forbidden for electric dipole atomic transitions due to the oscillator strength sum rule. In circuit QED, the capacitive coupling is analogous to the electric dipole one:  yet, such no-go property can be circumvented by Cooper pair boxes capacitively coupled to a resonator, due to their peculiar Hilbert space topology and a violation of the corresponding sum rule.
 \end{abstract} \maketitle (\today)
 \\

Cavity\cite{Raimond} and circuit\cite{Wallraff} quantum electrodynamics are fascinating topics due to the ultimate manipulation of light-matter interaction at the quantum level. In particular, the remarkable control of two-level systems
and their coupling to the quantum field of a resonator can pave the way to the realization and study of new quantum phases
for fundamental studies and quantum applications.
The physics of a system consisting of $N$ atoms coupled to the same quantum field of a photonic mode has generated great interest since
the celebrated paper by Dicke\cite{Dicke} on atomic superradiance. The Dicke model Hamiltonian is equivalent to a collection of
$1/2$-pseudospins (two-level systems) coupled to the same bosonic field. Long time ago, it was suggested that such a system
can undergo a {\it classical} phase transition\cite{Lieb}, i.e. occuring at a finite temperature. Remarkably, in such a phase transition,  the system exhibits a
spontaneous polarization of the atoms and a spontaneous coherent electromagnetic field. The Dicke model Hamiltonian exhibits also a {\it quantum} phase transition\cite{Brandes,Lehur}, which can occur at zero temperature by tuning the light-matter coupling across a quantum critical point\cite{QPT}.  Above the quantum critical point, the vacuum (ground state) of the cavity system is twice-degenerate. A linear superposition of these two degenerate ground states can be seen as a collective qubit characterized 
by a strong light-matter entanglement (the cavity field is in a coherent state, the matter part in a 'ferromagnetic' phase). 

In the case of the classical phase transition for  the Dicke model, such prediction was challenged by a no-go theorem\cite{pol1,pol2,pol3}. In fact, the Dicke Hamiltonian is obtained by neglecting the squared electromagnetic vector potential $\mathbf{\hat{A}}^2$ in the light-matter quantum Hamiltonian, but it was shown 
that by including such fundamental term in the Dicke Hamiltonian the finite-temperature classical phase transition disappears in the case of electric dipole coupling\cite{pol1,pol2,pol3,Knight}. Note that recent works in atomic cavity QED\cite{Dimer,esslinger}  have followed a dynamic approach to achieve a Dicke-like transition, i.e. by using time-dependent applied fields dressing the system. However, in order to have a genuine quantum phase transition and stable quantum phases associated to collective eigenstates, it is an important and fundamental matter to explore original approaches based on time-independent Hamiltonians in circuit QED. Indeed, a fundamental and general problem to  be investigated is whether intrinsic limitations occurring for cavity QED systems can translate necessarily into analogous constraints for superconducting circuit QED systems.  

In this paper, we show that in the case of cavity QED systems consisting of real atoms (coupled via their electric dipole) the due inclusion of the $\mathbf{\hat{A}}^2$-term in the Dicke Hamiltonian forbids the {\it quantum} phase transition as a consequence of the Thomas-Reiche-Kuhn (TRK) sum rule for the oscillator strength. Note that our proof is obtained by considering the time-independent quantum light-matter Hamiltonian. Remarkably, we  show that such no-go property does not necessarily hold for the analogous circuit QED system, namely a collection of Josephson atoms capacitively coupled to a transmission line resonator (see Fig. 1). We show that by taking as artificial atoms a collection of Cooper pair boxes the corresponding quantum Hamiltonian has the same form as the cavity QED counterpart, but, importantly, it has different constraints between the coupling constants. In the case of capacitive coupling, the peculiar topology of the Cooper pair box wavefunction is shown to be essential for the violation of the corresponding sum rule and the appearance of a quantum critical point in the case of capacitive coupling.
Our work shows that superconducting quantum circuits\cite{Nature_review} are not only appealing for the remarkable strength of the light-matter interaction, but they can give rise to artificial atoms with qualitatively different behavior when compared to real atoms. 

  \begin{center}
\begin{figure}[t!]
\includegraphics[width=260pt]{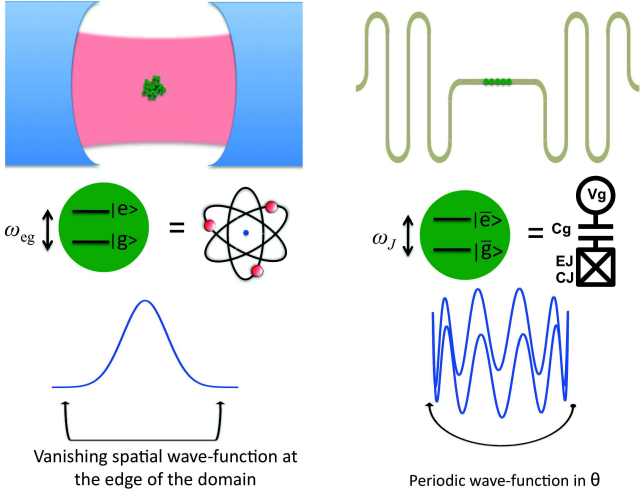}
\caption{
\label{sketch}
A sketch of the two considered systems. Left panel: a collection of $N$ identical atoms is coupled via their electric dipole
to the quantum field of a cavity resonator. Right panel: a collection of $N$ Cooper pair boxes is capacitively coupled to 
a transmission line resonator.}
\end{figure}
 \end{center}
\section{The case of cavity QED} 
Let us consider a collection of $N$ identical and isolated atoms, that do not interact directly with each other (see Fig. 1). This is equivalent to consider the Hamiltonian:
 \begin{equation}
 H = \sum_{j =1}^N H_j = \sum_{j=1}^{N}  \left ( \sum_{i_j=1}^{\nu} \frac{p_{i_j}^2}{2m_{i_j}^*} \right ) + U_j ,
 \end{equation}
 where $H_j$ is the bare Hamiltonian for the $j^{th}$ atom, $i_j$ denotes one of the $\nu$ particles of charge $q_{i_j}$ constituting the atom;  $\mathbf{p}_{i_j}\,$ ($m_{i_j}^*$)  is  the corresponding momentum (mass) entering the non-relativistic kinetic energy, while $U_j$ is the generic potential energy depending
on the position coordinates.  Of course, for atoms made of electrons we have $m_{i_j}^* = m_0$ and $q_{i_j}= -e$, but what follows is valid for the general case of different masses and charges. In order to account for the interaction of the atoms with the quantum field in the cavity resonator, we have to perform the standard minimal coupling replacement, namely 
$\mathbf{p}_{i_j}\rightarrow \mathbf{p}_{i_j}-q_{i_j} \hat{\mathbf{{A}}}(\mathbf{r}_{i_{j}})$. 
If we consider only one resonant photonic mode and suppose that in the region occupied by the atoms the spatial variation of the field is negligible, then we can replace $\hat{\mathbf{{A}}}(\mathbf{r})$ with $\mathbf{A_0} (a^{\dagger} + a) $, where $a^{\dagger}$ ($a$) is the corresponding bosonic photon creation (annihilation) operator and $\mathbf{A_0}$ is the vector potential field in the region occupied by the atoms. Therefore, we obtain the interaction term
\begin{equation}
H_{int}= - \sum_{j=1}^{N} \sum_{i_j=1}^{\nu} \frac{q_{i_j}}{m_{i_j}^*} \mathbf{p}_{i_j}\cdot \mathbf{A}_0 (a^{\dagger} + a) \end{equation} 
 as well as the  $\mathbf{A}^2$-term  
 \begin{equation}
 H_{A^2}=\sum_{j=1}^{N} \sum_{i_j=1}^{\nu} \frac{q_{i_j}^2}{2m_{i_j}^*}  \mathbf{A}_0^2 (a^{\dagger} + a)^2 . 
 \end{equation}
If we consider each atom as a two-level system ( $|\text{g}\rangle\,$ and  $|\text{e}\rangle\,$ are the two eigenstates for each atom) and if we use the fundamental commutation relation $i \hbar \frac{\mathbf{p}_{i_j}}{m_{i_j}^*}=[\mathbf{r}_{i_j},H_j]  $ , we get:
\begin{equation}
H_{int}=- i  \omega_{\text{eg}}\,\,\mathbf{d}_{\text{e}\,\text{g}} \cdot \mathbf{A}_0  (a^{\dag} + a) \sum_{j=1}^{N} \,(|\text{e} \rangle\,\langle \text{g}|)_j  \,\,+\,\, \rm{h.c.}\\ \label{pA}
\end{equation}
with $\omega_{\text{eg}}=\omega_{\text{e}}-\omega_{\text{g}}$ the atomic transition frequency and where the electric dipole matrix element is
$ \mathbf{d}_{\text{e}\,\text{g}}\, =\langle\text{e}|\sum_{i_j}^{\nu} q_{i_j} \mathbf{r}_{i_j}|\text{g} \rangle_j$.  It is convenient to introduce the bright excitation operators (bosonic in the 'thermodynamical' limit $N \gg 1$):
\begin{equation}
b^{\dag}\,=\,\frac{1}{\sqrt{N}}\,\sum_{j=1}^N(|\text{e} \rangle\,\langle \text{g}|)_j 
\end{equation}
in order to express the interaction Hamiltonian in the form:
 \begin{equation}
H_{int}=-i\hbar\,\Omega_0 (a+a^{\dag})\,b^{\dag}
 +\,\rm{h.c.}
\end{equation}
where\,
\begin{equation}
\Omega_0=\frac{\omega_{\text{eg}}}{\hbar} \,\mathbf{d}_{\text{e}\,\text{g}}\cdot \mathbf{A}_0 \sqrt{N\,}
\end{equation}
is the collective vacuum Rabi frequency accounting for the $\sqrt{N}$-enhancement of the coupling. 
The $\mathbf{\hat{A}}^2$-term takes instead the form
 \begin{equation}
 H_{A^2}=\,\sum_{i=1}^{\nu} \frac{q_i^2}{2 m_i^*}N \mathbf{A}_0^2 (a+a^{\dag})^2=\hbar D (a+a^{\dag})^2.
 \end{equation}
Finally, the atomic part of the Hamiltonian can be written as $H_{at}\,=\, \sum_{j\,}\,\hbar \omega_{\text{eg}} | \text{e} \rangle \langle \text{e}|_j = \hbar \omega_{\text{eg}} b^{\dag} b$, where we have omitted the bare energy of the dark excitations (the matter collective excitation modes not coupled to the cavity mode and orthogonal to the bright mode). Considering only one photonic mode, the cavity energy reads $H_{cav} = \hbar \omega_{cav} a^{\dag} a$.  By using an Hopfield-Bogoliubov transformation\cite{Hopfield,Ciuti_2005},
  it is possible to rewrite the Hamiltonian $H = H_{at} + H_{cav} + H_{int} + H_{A^2} $ in the form:
\begin{equation} 
H= \hbar \sum_{i=\pm} \omega_i P_i^ {\dag} P_i + E_G 
\end{equation}
where the  collective bosonic modes operators $P_+$ and $P_-$   satisfy $[P_+, P^{\dag}_+]=[P_-, P^{\dag}_-]=1\,\, \text{and}\,\,[P_+, P^{\dag}_-]=[P_+, P_-]=0 $. Such operators are defined as 
 $P_i = u^{ph}_i a + u^{el}_i b + v^{ph}_i a^{\dag} + v^{el}_i b ^{\dag}$, where the coefficient vectors $(u^{ph}_i, u^{el}_i, v^{ph}_i, v^{el}_i )^T$   are eigenvectors of the Hopfield-Bogoliubov matrix  $\mathcal{M}$:

  \begin{equation}
\label{HB}
  \mathcal{M}=\left (
  \begin{array}{cccc} \omega_{cav}+ 2D &  -i\Omega_0& -2D&-i\Omega_0\\i\Omega_0&  \omega_{\text{eg}}
  &  -i\Omega_0& 0 \\ 2D& -i\Omega_0 &-(\omega_{cav} + 2D) & -i\Omega_0 \\ -i\Omega_0& 0 & i\Omega_0& - \omega_{\text{eg}} \end{array}  \right).
\end{equation}
The corresponding positive eigenvalues directly give the frequencies $\omega_{+}$ and $\omega_-$ of the two collective modes.
In this formalism a quantum critical point is obtained when the determinant of $\mathcal{M}$ vanishes, because one excitation becomes gapless (zero eigenvalue).
Such determinant is simply given by:
\begin{equation}
\label{det}
Det(\mathcal{M}) = \omega_{\text{eg}} \omega_{cav} (\omega_{\text{eg}} (4 D+ \omega_{cav}) - 4 \Omega_0^2) .
\end{equation}
Note that by using a Holstein-Primakoff approach\cite{Brandes} (details not shown) one finds the same equation $Det(\mathcal{M}) = 0$ for the quantum critical point.
In the case of the Dicke Hamiltonian, the $\mathbf{\hat{A}}^2$-term is neglected, so $D = 0$ and the condition $Det(\mathcal{M}) = 0$ gives the quantum critical coupling 
$\Omega_{c}^{Dicke} = \sqrt{\omega_{\text{eg}} \omega_{cav}}/2$. Above a critical coupling, a symmetry breaking occurs, the ground state (vacuum) becomes twice degenerate, a spontaneous coherence for light and matter fields appear (the bosonic excitations have to be considered around the symmetry breaking vacua, e.g., via an Holstein-Primakoff approach \cite{Brandes}). 
In Fig. 2a, the excitation frequencies are shown for the Dicke case $D=0$ (taking $\omega_{cav} = \omega_{eg}$). The $\omega_-$ branch vanishes at the quantum critical point. However, if the $\mathbf{\hat{A}}^2$-term is retained, the situation changes dramatically.
In fact, by using the Schwartz inequality, we get :
\begin{equation}
\label{Scw}
    \Omega_0^2= \frac{\omega_{\text{eg}}^2}{\hbar^2} N |\mathbf{d}_{\text{e}\,\text{g}} \cdot \mathbf{\mathbf{A}_0}|^2   \leq \frac{\omega_{\text{eg}}^2}{\hbar^2}  N |\mathbf{d}_{\text{e}\,\text{g}}|^2  | \mathbf{\mathbf{A}_0}|^2. \end{equation}
 By using the Thomas-Reiche-Kuhn (TRK) sum rule for the electric dipole oscillator strength (for details see Methods) we get:
 \begin{equation}
 \label{ineq}
  D \geq   \frac{\Omega_0^2}{ \omega_{\text{eg}}}.
 \end{equation}
The general inequality $D \geq \Omega_0^2/\omega_{\text{eg}}$  implies that   $Det(\mathcal{M})$ {\it never} vanishes (as it can be verified straightforwardly by inspecting the expression in  Eq. (\ref{det})) and that the quantum critical point does not exist any longer (see Fig. 2c and d). 
 It is interesting to note that the last inequality also prevents a classical superradiant phase transition \cite{pol1,pol3}. 
In conclusion, under rather general hypothesis,  we have proved that the quantum critical point disappears in presence of the $\mathbf{\hat{A}}^2$-term (for electric dipole transitions and if no external optical fields are applied to the system\cite{time}). The reason is that, as shown by Eq. (\ref{ineq}), the amplitude of the $\mathbf{\hat{A}}^2$-term is not independent of the vacuum Rabi coupling. On the contrary, the  $\mathbf{\hat{A}}^2$-term becomes even dominant in the ultrastrong coupling regime ($\Omega_0/\omega_{\text{eg}} \gg 1$ implies $D \gg \Omega_0$). We would like to point out that such no-go proof does not concern the case of magnetic coupling (i.e., with an interaction Hamiltonian $\vec{\mu} \cdot \vec{B}$ between real spins and a quantized magnetic field).

 \section{The case of circuit QED}
 
Here, we show why the fundamental limitation for the superradiant quantum phase transitions, which we have proved under rather general assumptions for the case of cavity QED based on electrical dipole coupling, does not generally apply to superconducting circuit QED. In the case of superconducting Josephson atoms, $\hat{\theta}$ and $\hat{n}$ (phase and number operator respectively) play a role analogous to electron spatial position and momentum operators for a real atom. When a Josephson junction is embedded in a transmission line resonator, its degrees of freedom are coupled to the bosonic quantum field of the resonator. In circuit-QED, there are many types of artificial atoms depending on the connection of the Josephson junctions to capacitors and inductors. The analogous of the electric dipole coupling is given by the so-called capacitive coupling, while the magnetic coupling in cavity QED corresponds to the inductive coupling\cite{Devoret_strong,degvacua}. The quantum phase transitions of a chain of inductively coupled Josephson atoms was recently studied in Ref. \cite{degvacua}. Here we will treat the important case of capacitive coupling in circuit QED. As artificial atom, we will consider a Cooper pair box and show that the peculiar wavefunction topology of such a system can lead to the appearance of a superradiant quantum critical point due to a violation of the sum rule.
 
\subsection{A chain of Cooper pair boxes capacitively coupled to a transmission line resonator}
Let us consider a chain of Cooper pair boxes embedded in a transmission line resonator\cite{blais}. 
For simplicity, we consider that the Josephson atoms are placed in the middle of the resonator 
in a small region where the spatial variation of the resonator field is negligible.
Moreover we consider only one resonator mode. With these assumptions (analogous to what was done in the case of cavity QED for real atoms), we have $\hat{V} = \mathcal{V}(\alpha + \alpha^{\dag})$   where $\alpha$  and $\alpha^{\dag}$ define the bosonic annihilation and creation operators of the resonator field. The quantum Hamiltonian then reads
\begin{widetext}
\begin{equation}
H = \hbar \omega_{\text{res}} \alpha^{\dag} \alpha  + \sum_{j=1}^{N}  \left \{ 4 E_c \sum_{n \in \, \mathbb{Z}}
(n -(\hat{n}_{ext})_{j})^2 |n\rangle \langle
n|_{j} \nonumber 
- \frac{E_{J}}{2}\sum_{n \in\, \mathbb{Z}}(|n+1\rangle\langle n|+ |n\rangle \langle
n+1|)_{j} \right \}
\end{equation}
\end{widetext}

with $E_c=\frac{e^2}{2(C_J+C_g)}$ the charging energy, $E_{J}$ the tunable Josephson energy. Finally, $(\hat{n}_{ext})_{j}=\frac{C_g}{2e} (V_g+\hat{V})_{j}$ is the excess charge in the $j$-th Cooper pair box depending on the static control gate voltage $(V_g)_{j}$ and on the quantum voltage $\hat{V}$.
\begin{widetext}
\begin{center}
\begin{figure}[t!]
\includegraphics[width=450pt]{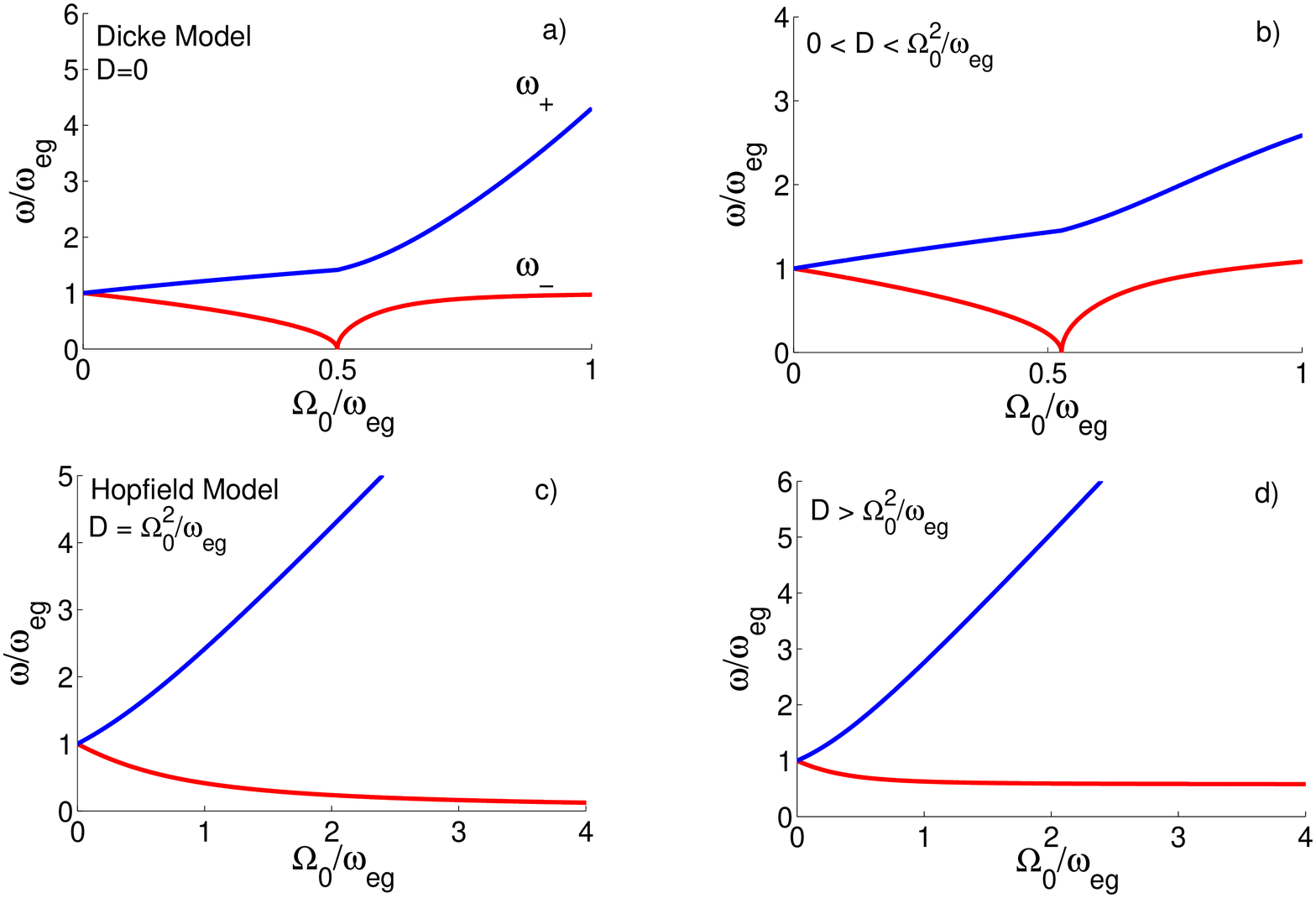}
\caption{Normalized frequencies of the bosonic collective excitations $\omega_{+}$ (blue line, upper branch) and $\omega_{-}$ (red line, lower branch) as a function of the normalized vacuum Rabi frequency $\Omega_0$. Note that we have considered the case of real (artificial) two-level atoms with a transition frequency equal to the 
cavity (resonator) frequency. For the two considered systems (see Fig. \ref{sketch}), the form of the quantum Hamiltonian is identical and depends on four physical quantities:
the resonator mode frequency; the two-level transition frequency $\omega_{eg}$; the collective vacuum Rabi frequency $\Omega_0$; the amplitude frequency $D$ of the Hamiltonian term quadratic in the resonator boson field. In the case of cavity QED, such a term is due to the squared electromagnetic vector potential. In the considered class of circuit QED systems (Cooper pair boxes capacitively coupled to a resonator), it is due to the squared voltage field operator. The frequencies $\omega_{+}$ and $\omega_{-}$ are very sensitive to the relation between $\Omega_0$ and $D$. There are 4 possible cases:
(a) $D=0$ (no term quadratic in the resonator quantum field).  This case corresponds to the Dicke model. A quantum critical point (the excitation frequency $\omega_{-}$ vanishes at that point) is present for a dimensionless vacuum Rabi frequency equal to $0.5$ (the energies above the critical point have been calculated through an Holstein-Primakoff approach\cite{Brandes}). According to our proof, this limit cannot be obtained in the case of cavity QED with atoms coupled via electric dipole (in the absence of time-dependent applied fields).  (b) $0 < D < \Omega_0^2/\omega_{eg}$. In this case, the superradiant quantum phase transition is still possible, but with a shifted quantum critical point. This case is still not accessible to the considered cavity QED system, but it can be achieved with Cooper pair boxes in circuit QED.  In the simulation, $D= 0.1 \,\Omega_0^2/\omega_{eg}$ (it can be obtained with Cooper pair boxes such that $E_J/(4E_C) = 0.1 $). (c) $D=  \Omega_0^2/\omega_{eg}$. The quantum critical point disappears and $\omega_- \to 0$ for $\Omega_0 \to + \infty$. This corresponds to the Hopfield model \cite{Ciuti_2005}. It can be obtained in cavity QED if the electric dipole transition takes all the oscillator strength. (d) $D >  \Omega_0^2/\omega_{eg}$. 
No quantum phase transition and the frequency $\omega_-$ is finite even in the ultrastrong coupling limit. }
\end{figure}
 \end{center}
 \end{widetext}
 
Here, we will consider the ideal situation of  identical Cooper pair boxes such as
$n_g^j =\frac{C_g}{2e} (V_g)_{j}= \frac{1}{2}$. In each artificial atom the two-level
subspace is obtained by keeping only the two
first number states $|n=0\rangle$\,and
$|n=1\rangle$. In such a subspace, for $n_g = 1/2$,  the eigenstates
of a Cooper pair box due to the Josephson coupling are
$ |\bar{g} \rangle=\frac{1}{\sqrt{2}}(|n=0\rangle +
|n= 1\rangle)\,$ and $ |\bar{e} \rangle=\frac{i}{\sqrt{2}}(|n=0\rangle -
|n=1\rangle)\,$, while  the transition energy is $E_J= \hbar
\omega_{J}$. Since $(\hat{n}-\hat{n}_{ext})^2 =  (\hat{n}-n_g)^2 - 2 \frac{C_g}{2e} (\hat{n}-n_g) \hat{V} + ( \frac{C_g}{2e}  \hat{V} )^2$, it is clear that $(\hat{n}-n_g)$ is analogous to the electron momentum, $(\hat{n}-n_g) \hat{V}$ is analogous to the $\mathbf{p}\cdot \mathbf{\hat{A}}$ coupling term, while $ \hat{V}^2$ is reminiscent of the $\mathbf{\hat{A}}^2$ term. The interaction between the resonator quantum field and the charge of the Cooper pair boxes read:
\begin{equation}
H_{coupl} = -i  4 E_c \frac{C_g}{2e}  \mathcal{V} (\alpha + \alpha^{\dag}) \sum_{j=1}^{N} \,(|\bar{\text{e}} \rangle\,\langle \bar{\text{g}}|)_j  \,\,+\,\, \rm{h.c.} 
\end{equation}
which is analogous to the interaction term in Eq. (\ref{pA}).
The Hamiltonian contribution due to the $n_{ext}^2$ term reads:
\begin{equation}
H_{V^2} = \sum_{j=1}^N 4 E_c (\frac{C_g}{2e})^2 {\mathcal V}^2 (\alpha + \alpha^{\dag})^2 = \hbar \bar{D} (\alpha + \alpha^{\dag})^2 .
\end{equation}
As in the case of cavity QED, we introduce the operator
\begin{equation}
 {\beta}^{\dag}\,=\,\frac{1}{\sqrt{N}}\,\sum_{j=1}^N(|\bar{\text{e}} \rangle\,\langle \bar{\text{g}}|)_j ,
 \end{equation}  
 and we have
 \begin{equation}
H_{coupl} =-i \hbar \bar{\Omega}_0 (\alpha + \alpha^{\dag}) \beta^{\dag} \,\,+\,\, \rm{h.c.} 
\end{equation}
 with 
 \begin{equation}
 \bar{\Omega}_0 = \frac{4 E_c}{\hbar} (\frac{C_g}{2e})\sqrt {N} {\mathcal V}.
 \end{equation}
 The sum of the bare energies of the Cooper pair boxes reads
 $H_{CPB} = \hbar \omega_J \beta^{\dag} \beta $ 
(as in the case of real atoms, we have omitted the dark matter excitations, orthogonal to the bright mode
and uncoupled to the resonator mode). 
  To determine the excitation bosonic modes of
  the total quadratic Hamiltonian $\bar{H} = H_{\text{res}} + H_{CPB} + H_{coupl} + H_{V^2}$ , one can use the Hopfield-Bogoliubov transformation 
  for which one has to diagonalize  the matrix $\bar{\mathcal{M}}$ 
\begin{equation}
\label{HB}
 \bar{\mathcal{M}}=\left (
  \begin{array}{cccc} \omega_{res}+ 2\bar{D} & -i \bar{ \Omega}_0& -2\bar{D}&-i \bar{\Omega}_0\\ i \bar{ \Omega}_0&  \omega_{J}
  &  -i\bar{\Omega}_0& 0 \\ 2\bar{D}& -i \bar{ \Omega}_0&-(\omega_{res} + 2\bar{D}) & - i \bar{ \Omega}_0 \\  -i \bar{ \Omega}_0& 0 & i\bar{ \Omega}_0& - \omega_{J} \end{array}  \right).
\end{equation}
By looking at the matrix in Eq. (\ref{HB}), it is clear that in the considered circuit QED system we have found an Hamiltonian of the same form as the one obtained for the cavity QED case. However, there is a fundamental difference, which is crucial for the existence of a superradiant quantum critical point: the relation between the vacuum Rabi frequency $\bar{ \Omega}_0$ and the term $\bar{D}$ originating from the squared of the resonator field. In fact, we have
\begin{equation}
\bar{D} = \frac{\bar{ \Omega}_0^2}{\omega_J}  \frac{E_J }{4 E_c}.
\end{equation}
Hence, in contrast to the cavity QED case, $\bar{D}$ can be made smaller than $\frac{\bar{ \Omega}_0^2}{\omega_J}$.
In particular if $\frac{E_J }{4 E_c} \ll  1$ then $0 < \bar{D} \ll \frac{\bar{ \Omega}_0^2}{\omega_J} $. In such a case, the quantum critical
point is shifted by the finite $D$ term, but it still exists (see Fig. 2b).   Indeed, the equation $det( \bar{\mathcal{M}}) = 0$ gives the quantum critical coupling:
\begin{equation}
\bar{\Omega}^c = \frac{ \sqrt{\omega_{res} \omega_{J}}}{2 \sqrt{1
-\frac{E_J}{4E_c}}}.
\end{equation}
Note that the limit $\frac{E_J }{4 E_c} \ll 1$ is also the one where the two-level approximation for the Cooper pair box is excellent.
\subsection{The role of the Cooper pair box wavefunction topology}

What is the difference with respect to the cavity QED atomic case ? 
The key issue here is that the wavefunction of a Cooper pair  box has a different topology when compared to the wavefunction of a real atom. In the case of a real atom, the wavefunction is not periodic. 
The commutator between position and momentum is proportional to the identity operator, namely $[x, p_x] = i \hbar$.
Instead, in the case of a Cooper pair  box, the charge is quantized and correspondingly the wavefunction is periodic with respect to the phase $\theta$ , the phase difference across the junction. In the phase representation, the Cooper pair box wavefunction is such that $\Psi_{CPB}(\theta + 2 \pi) = \Psi_{CPB}(\theta )$. 
In other words, there is a circular topology that changes the algebraic rules of the Hilbert space where the physical states 'live'. In particular, the commutator between the phase operator $\hat{\theta}$ (analogous to the position) and the number operator $\hat{n}$ (analogous to the momentum) is {\it not always} proportional to the identity operator. This is crucial because the TRK sum rule for the atomic oscillator strength is due to the fact that $[x, p_x] $ is proportional to the identity. The TRK sum rule is responsible for the inequality $D \geq \Omega_{0}^2/\omega_{eg}$ in the case of real atoms. In the case of Cooper pair  boxes,
we have indeed shown that by choosing a large capacitance energy $E_C$  it is possible to strongly violate the sum rule  ($\bar{D} \ll \bar{\Omega}_{0}^2/\omega_{J}$)   and therefore to allow the existence of a quantum critical point.
Note that the violation of the TRK sum rule has been already studied in the context of quantum rigid rotators \cite{trkrotator}, which have a different topology with respect to real atoms, but similar to the Cooper pair box. Here, the fundamental topological properties of the Cooper pair box have very important consequences on the collective behavior of the chain in a resonator and on the existence of a quantum critical point for the case of capacitive coupling.

\section{Comparison with dynamical approaches for the Dicke transition}
In the considered circuit-QED systems the corresponding quantum Hamiltonians are time-independent (no applied pump fields dressing the system). The predicted quantum phase transition occurs at zero temperature and affects the quantum ground state (vacuum) subspace. Above the quantum critical point, there are two degenerate ground states $\vert G_{\pm} \rangle = \vert\pm\alpha \rangle  \prod_{j=1}^N \vert \pm \rangle_j  $, where $\vert \alpha \rangle$ is a coherent state for the cavity mode and $\vert \pm \rangle_j = \frac{1}{\sqrt{2}} (\vert e\rangle_j \pm \vert g \rangle_j)$. In principle one could encode quantum information in linear superpositions of such collective vacua, which have been shown to enjoy some protection with respect to local noise sources\cite{degvacua}. It is important to point out that in atomic cavity QED systems interesting dynamical approaches to achieve a superradiant transition have been proposed\cite{Dimer} by using applied laser fields and a recent impressive experimental demonstration has been reported \cite{esslinger} (by using a coherently dressed Bose-Einstein condensate). In the dynamical case, observables can indeed show interesting discontinuities and symmetry breaking signatures around the critical coupling. However, with respect to the time-independent case, in the dynamic configuration the system continuously emits (Raman) photons. In fact, in a dynamical approach with time-dependent pump fields, one can obtain a Dicke Hamiltonian only in the pump rotating frame. As clearly proven in the general paper by Brown\cite{Brown}, in such a rotating frame, the Hamiltonian describing the coupling of the system to the environment (e.g., the extracavity electromagnetic field) becomes time-dependent and has an impact on the coherence properties. Namely, in the case of an open system with dressing time-dependent fields, the Hamiltonian $H_{syst}(t)$ does depend on time. If the dressing fields are monochromatic (e.g., a pump field with a time-dependance as $e^{-i \omega_p t}$), one can conveniently pass to the so-called rotating frame to eliminate such time-dependence. In such a frame, the Hamiltonian becomes $\tilde{H}_{syst}= U(t) H_{syst}(t) U^{\dagger}(t)$ where $U(t)$ is the unitary transformation connecting the laboratory frame to the rotating frame. In the rotating frame the Hamiltonian $\tilde{H}_{syst}$ does not depend on time and one can define pseudo-eigenstates such that $ \tilde{H}_{syst}  \vert\tilde{ \Psi}_j \rangle = \tilde{E}_j \vert \tilde{\Psi}_j \rangle$. In a dynamical approach for the superradiant transition, $\tilde{H}_{syst}$ is the Dicke Hamiltonian. However, the Hamiltonian term describing the interaction with the reservoir becomes time-dependent. Namely, one finds $\tilde{H}_{syst-bath}(t) = U(t) H_{syst-bath} U^{\dagger}(t)$.  Therefore, if one prepares the system in the rotating frame Dicke pseudo-ground state $\vert\tilde{ \Psi}_0 \rangle$, then such a state will not be stationary:  in particular,  the reservoir Hamiltonian $\tilde{H}_{syst-bath}(t)$ can induce transitions from the pseudo-ground state $\vert\tilde{ \Psi}_0 \rangle$ to excited pseudo-eigenstates even if the reservoir is at zero temperature. In the laboratory frame, this corresponds to the continuous emission of photons by the system. 
Consequently, the steady-state of the open system in the dynamical configuration is  not a pure eigenstate of the Dicke Hamiltonian $\tilde{H}_{syst}$, but a more complex non-equilibrium state. In particular, it is not possible to have stable superpositions between the two degenerate ground states of the Dicke Hamiltonian. Hence, we believe that circuit-QED systems with time-independent Hamiltonians have a considerable interest in this respect.

 \section{conclusion}
In conclusion, we have proved that, under rather general assumptions (electric dipole coupling, two-level  atoms, single-mode resonator field), for the case of real atoms in a cavity the superradiant Dicke quantum phase transition disappears if the $\mathbf{\hat{A}}^2$-term is correctly included in the quantum light-matter Hamiltonian. We have shown that the oscillator strength sum rule is what forbids the existence of a quantum critical point. For comparison, we have considered the circuit QED system consisting of a collection of Josephson atoms capacitively coupled to a a transmission line resonator, being the capacitive coupling in circuit QED the analogous of electric dipole coupling in cavity QED. We have shown that it is possible to obtain the same quantum Hamiltonian as for the cavity QED case. However, by considering as artificial Josephson atom a Cooper pair box, we have shown that it is possible to have a quantum critical point. In the case of capacitive coupling, the peculiar circular topology of the Cooper pair box wavefunction prevents the same fundamental limitation imposed by the oscillator strength rule.  Our theoretical work shows that circuit QED can provide access to interesting physical effects which are not necessarily obtained in the analogous cavity QED system, in spite of formal analogies. These original properties may pave way to exciting avenues towards the realization of interesting quantum phases for fundamental studies and quantum applications.  

We would like to thank I. Carusotto, M.H. Devoret, S. De Liberato for discussions.

 \section{Methods}
 Here we show in detail why the TRK sum rule leads to the inequality in Eq. (\ref{ineq}). 
 Let us call  $|\text{g} \rangle_j$ and $ |\text{e} \rangle_j$,  the two eigenstates of the $j$-th atom. Using the fundamental commutation relation $i\hbar \frac{\mathbf{p}_{i_j}}{m_{i_j}^*}=[\mathbf{r}_{i_j},H_j]  $, we have  for $ |\sigma\rangle |\sigma'\rangle_j \in  \{ |\text{g} \rangle_j , |\text{e} \rangle_j \} $, 
\begin{eqnarray}
\langle \sigma | \sum_{i_j=1}^{\nu} \frac{q_{i_j}}{m_{i_j}^*} \mathbf{p}_{i_j}  |\sigma'\rangle_j
=-\frac{i}{\hbar}  \langle  \sigma | [ \mathbf{d}_{j},H_j] |\sigma'\rangle_j \\\nonumber
\end{eqnarray}
where we introduced the dipole operator $\mathbf{d}_{j} = \sum_{i_j=1}^{\nu} q_{i_j}\mathbf{r}_{i_j}$.
Since all the atoms are identical, we can omit the index
 $j$ to simplify the notations:
\begin{eqnarray}
\langle \sigma| \sum_{i=1}^{\nu} \frac{q_{i}}{m_{i}^*} \mathbf{p}_{i}  |\sigma'\rangle
=-i(\omega_{\sigma'}-\omega_{\sigma})\langle  \sigma |  \mathbf{d}|\sigma'\rangle 
 \nonumber \\-i \sum_{i=1}^{\nu} \frac{1}{2m_{i}^*}\int\limits_V  \psi_{\sigma}^* (\mathbf{r}_{i_1},.,\mathbf{r}_{i_{\nu}}) \frac{\partial^2}{\partial^2 \mathbf{r}_{i}} ( \mathbf{d}\, \psi_{\sigma'}(\mathbf{r}_{i_1},.,\mathbf{r}_{i_{\nu}})) d^{\nu} \mathbf{r}  \nonumber\\
+i \sum_{i=1}^{\nu} \frac{1}{2m_{i}^*}\int\limits_V  \mathbf{d}\, \psi_{\sigma'}(\mathbf{r}_{i_1},.,\mathbf{r}_{i_{\nu}}) \frac{\partial^2}{\partial^2 \mathbf{r}_{i}}  (\psi_{\sigma}^* (\mathbf{r}_{i_1},.,\mathbf{r}_{i_{\nu}}) ) d^{\nu} \mathbf{r}
\end{eqnarray}
We then define $ \mathbf{d}_{\text{e}\,\text{g}} = \langle  \text{e} |  \mathbf{d}|\text{g}\rangle =  | \mathbf{d}_{\text{e}\,\text{g}}| \mathbf{u}  $ with $\mathbf{u}$ an unitary vector.\\
In the last derivation,  on purpose we made explicit two integrals whose difference gives zero with the sufficient condition of vanishing  wave-functions (and their first and second derivatives) at the edge of the quantization box volume $V$, as we can see after a double integration by parts.
Such boundary conditions are of course valid in the case of atomic states since their spatial wave-functions, and all their derivatives vanish at the infinity. This is not valid in general for Cooper pair boxes, whose  eigenstates have wavefunctions $2\pi$-periodical in $\theta$, the phase difference across the junction. Indeed,  the Cooper pair box eigenstates are linear superpositions of quantized charge states $\{ |n\rangle \}_{n \in\, \mathbb{Z}} $ whose wavefunctions are  $\Psi_n(\theta) = \frac{1}{\sqrt{2\pi}} e^{i n \theta}$.\\
Now, to get  the inequality (\ref{ineq}), we write:
 $$\sum_{1}^{\nu} \frac{q_i^2}{2 m_i^*}=\frac{-i}{2\hbar}  \langle \text{g} |[\sum_{i=1}^{\nu}  q_i \mathbf{r}_i.\mathbf{u} , \sum_{i=1}^{\nu} \frac{q_i}{m_i^*} \mathbf{p}_{i}.\mathbf{u^{\star}}] | \text{g} \rangle  $$
Using the completeness relation $\sum_{\sigma} | \sigma\rangle \langle \sigma| = 1$,  $ | \sigma\rangle$  (denoting the generic atomic eigenstate), and using $i\hbar \frac{\mathbf{p}_{i}}{ m_i^*}=[\mathbf{r}_i,H]$, we get:
 \begin{eqnarray}
  \begin{array}{ccc} 
 \sum_{1}^{\nu} \frac{q_i^2}{2 m_i^*} =  \frac{ 1 }{\hbar}\sum_{\sigma}(\omega_{\sigma}-\omega_{\text{g}})|\langle \text{g} |\sum_{i=1}^{\nu}  q_i \mathbf{r}_i.\mathbf{u} |\sigma\rangle | ^2\\\nonumber
    \geq  \frac{1}{\hbar} \omega_{\text{eg}} |\langle \text{g} |  \sum_{i=1}^{\nu} q_i \mathbf{r}_i.\mathbf{u} |\text{e}\rangle | ^2 =   \frac{1}{\hbar} \omega_{\text{eg}}   | \mathbf{d}_{\text{e}\,\text{g}}|^2
   \end{array} 
 \end{eqnarray}
 where the quantity $\frac{ 1}{\hbar} (\omega_{\sigma}-\omega_{\text{g}})\vert \mathbf{d}_{ \text{g}, \sigma}\cdot\mathbf{u} \vert^2 $ is 
 proportional to the oscillator strength $f^{\mathbf{u}}_{\text{g},\sigma}$
 of the atomic transition between $\vert \text{g} \rangle$ and  $\vert \sigma \rangle$ in the direction $\mathbf{u} $ . If this oscillator strength sum rule is applied to Eq. (\ref{Scw}), one finally obtains the inequality (\ref{ineq}).
For the Cooper pair box, due to the circular topology of the wavefunction, the commutator between charge and phase operator is not always proportional to the identity and the sum rule does not hold.

\end{document}